\documentclass[12pt]{article}
\usepackage{amsmath}
\usepackage{amssymb}
\usepackage{graphicx}

\begin{document}

\begin{center}
\smallskip \ 

\textbf{A SPINNING PARTICLE IN A M\"{O}BIUS STRIP}

\smallskip \ 

\smallskip \ 

\smallskip \ 

J. A. Nieto$^{\star \dagger }$ \footnote{%
nieto@uas.uasnet.mx} and R. P\'{e}rez-Enr\'{\i}quez$^{^{\ast }}$ \footnote{%
rpereze@correo.fisica.uson.mx}

$^{\star }$\textit{Facultad de Ciencias F\'{\i}sico-Matem\'{a}ticas de la
Universidad Aut\'{o}noma} \textit{de Sinaloa, 80010, Culiac\'{a}n, Sinaloa, M%
\'{e}xico.}

\smallskip \ 

$^{\dagger }$\textit{Departamento de Investigaci\'{o}n en F\'{\i}sica de la
Universidad de Sonora, 83000, Hermosillo Sonora, M\'{e}xico}

\smallskip \ 

$^{\ast }$\textit{Departamento de F\'{\i}sica de la Universidad de Sonora,
83000, Hermosillo, Sonora , M\'{e}xico}

\bigskip \ 

\bigskip \ 

\textbf{Abstract}
\end{center}

We develop the classical and quantum theory of a spinning particle moving in
a M\"{o}bius strip. We first propose a Lagrangian for such a system and then
we proceed to quantize the system via the constraint Hamiltonian system
formalism. Our results may be of particular interest in several physical
scenarios, including solid state physics and optics. In fact, the present
work may shed some new light on the recent discoveries on condensed matter
concerning topological insulators.

\bigskip \ 

\bigskip \ 

\bigskip \ 



February, 2011

\newpage \noindent \textbf{1. Introduction}

\smallskip \ 

Recent discoveries on condensed matter physics have opened new ways on
topology and physics. In particular, this is evident in the new state of
matter known as \textit{topological insulators} which behaves as insulators
in the bulk but have topological stable electronic states at its border;
giving rise to a conductor state [1-2]. The explanation of this phenomenon
resides on the type of electronic states that are present at the border:
they are topologically different to those states in the bulk.

A way to interpret this condition of matter has been proposed as the case of
a cylinder and a M\"{o}bius strip [3]. Both surfaces can be obtained from
the gluing the two extremes of a rectangle band: one directly but the other
after a half twist. In such a case, it is impossible to transform one into
the other. In fact, the M\"{o}bius strip of the latter example cannot be
modified in any way onto the cylinder of the former. As Zhang [4] has
remarked this phenomena is the kind of difference between the electronic
states in the topological insulator. The electronic states in the bulk have
an energy gap at the Fermi level but at the border, the electronic states
lay within the gap giving the material its metallic properties.

The existence of this type of behavior was predicted in 2006 for
bi-dimensional systems with band structure in which quantum spin Hall effect
arises from large spin orbit interaction. Zhang and his collaborators,
proposed HgTe quantum dots as potential candidates and one year later, in
2007, they were reporting the first realization of these materials [5-6].

Subsequent studies [7-8] have found materials such as the $Bi_{2}Te_{3}$, in
which a three-dimensional topological insulator state has been observed.
Moreover, following the M\"{o}bius - cylinder parallelism, some researchers
have studied graphene type structures with one half twist, two and several
more twist, observing whether they present the kind of edge states which we
mentioned above. Two years ago, in 2009, Z. L. Guo \textit{et al}. published
a Physical Review B paper in which they analyze the edge states in a M\"{o}%
bius graphene strip [9]. In fact, they describe a current through a M\"{o}%
bius ring. Moreover, last year, Wang and colaborators, made some \textit{ab
initio} studies for M\"{o}bius graphene strips with fixed length but
changing its width [10]. They observed the edge states with non zero
magnetic moment, confirming the topological insulator behavior.

Even more recently Chih-Wei Chang \textit{et al.} [11] have discovered that
electromagnetic M\"{o}bius symmetry can be successfully introduced into
composite metamolecular systems made from metals and dielectrics. This
discovery opens the door for exploiting novel phenomena in metamaterials.

In all these studies the starting point is the Schr\"{o}dinger equation with
appropriate boundary conditions. However, the results of this studies depend
very much on the model involved. This has to do with the lack of a
systematic way to consider the M\"{o}bius strip at the classical and quantum
levels. Motivated by these observations in this work we develop the
classical and quantum theory of a spinning object moving in a M\"{o}bius
strip. We first propose a Lagrangian for such a system and then we proceed
to quantize the system via the Dirac's constraint Hamiltonian system
formalism [12]-[13] (in particular see Ref. [14] and references therein). We
prove that this Lagrangian-Hamiltonian M\"{o}bius strip mechanism can be
used in general and not only in any particular case.

The paper is organized as follows. In the section 2, we briefly describe the
topology of M\"{o}bius strip and discuss some of its special properties. In
section 3, we present a brief description of the Hamiltonian constraint
method. In section 4, we discuss the quantization of the classical motion of
a spinning particle in the M\"{o}bius strip. In last section, we discus some
of the characteristics of those states and extract some conclusions.

\bigskip \ 

\noindent \textbf{2. M\"{o}bius Strip}

\smallskip \ 

The M\"{o}bius strip [15-16] is a bi-dimensional manifold with only one
face, and can be built from a strip of paper as follows: took the band and
join together its both ends but being sure to twist one of them a half turn
as it is shown in Figure 1. It is possible to confirm that the M\"{o}bius
strip has one side and a single border.

\begin{figure}[htbp]
\centering
\includegraphics[scale=0.3,keepaspectratio=true]{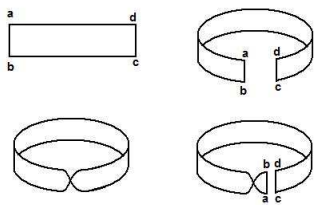}
\caption{{\protect \footnotesize How to build a M\"{o}bius strip }}
\end{figure}

We observe that the vector $\hat{n}$, perpendicular to the plane at any
point $\bar{p}$ on the M\"{o}bius strip, will change its direction as we
move along the centre line(meridian) of the strip; this normal vector will
be $-\hat{n}$ when we return to the $\bar{p}\ $point. This means that our
strip cannot be oriented in space and for this reason we say that the M\"{o}%
bius strip is a non-oriented surface.

Indeed, the presence of a M\"{o}bius strip can used to define the concept of
oriented surface; any surface containing a band topologically equal to a M%
\"{o}bius strip it is a non-oriented manifold [15]. Another way to say this,
is that a surface is oriented if it has two sides.

A straight forward result that can be reached from the definition of this
interesting strip, is that the normal vector $\hat{n}\ $at the meridian
requires to go through the circle twice in order to return to its original
orientation; such a behavior is similar to that of the spin in the
Stern-Gerlach experiments discussed by Feynmann [17].

If we want to analyses the movement of a spinning particle moving along a M%
\"{o}bius strip, we need to consider some relevant facts on this
non-oriented surface. First of all, we look at the parametric representation
of this surface:

\begin{equation}
\begin{array}{c}
x=R\sin u+v\sin {\frac{u}{2}}\sin u, \\ 
\\ 
y=R\cos u+v\sin {\frac{u}{2}}\cos u, \\ 
\\ 
z=v\cos {\frac{u}{2},}%
\end{array}
\tag{1}
\end{equation}%
where $u\in \lbrack 0,2\pi )\ $ and $v\in \lbrack -1/3,+1/3]$. As we can
see, if the particle is moving along de meridian line i.e. when $v=0$; it
describes a circle of radius $R$ in the plane $xy$. However, as the particle
moves around this circle, the vector normal to the surface turns in such a
way that it needs two complete turns to recover its original orientation.
This effect can be seen when one translate the normal vector $\hat{n}$ along
the meridian.

Consider the directional derivatives $e_{(A)}^{i}=\frac{\partial x^{i}}{%
\partial \xi ^{A}}$, with $\xi ^{A}=(u,v)$. Using (1) we get

\begin{equation}
\begin{array}{l}
e_{1}^{1}=\frac{\partial x}{\partial u}=R\cos u+v\left( \frac{1}{2}\cos {%
\frac{u}{2}}\sin u+\sin {\frac{u}{2}}\cos u\right) , \\ 
\\ 
e_{1}^{2}=\frac{\partial y}{\partial u}=-R\sin u+v\left( \frac{1}{2}\cos {%
\frac{u}{2}}\cos u-\sin {\frac{u}{2}}\sin u\right) , \\ 
\\ 
e_{1}^{3}=\frac{\partial z}{\partial u}=-\frac{1}{2}v\sin {\frac{u}{2},} \\ 
\\ 
e_{2}^{1}=\frac{\partial x}{\partial v}={\sin {\frac{u}{2}}\sin u,} \\ 
\\ 
e_{2}^{2}=\frac{\partial y}{\partial v}={\sin {\frac{u}{2}}\cos u,} \\ 
\\ 
e_{2}^{3}=\frac{\partial z}{\partial v}={\cos {\frac{u}{2}.}}%
\end{array}
\tag{2}
\end{equation}

Note that when $v=0$, from (1) one finds that the only nonvanishing
components of $e_{(A)}^{i}$ are%
\begin{equation}
\begin{array}{l}
e_{1}^{1}=\frac{\partial x}{\partial u}=R\cos u, \\ 
\\ 
e_{1}^{2}=\frac{\partial y}{\partial u}=-R\sin u,%
\end{array}
\tag{3}
\end{equation}%
which correspond to a circle of radius $R$ in the $xy$-plane.

The normal vector $\hat{n}$ at the centre line is given by

\begin{equation*}
\hat{n}=\frac{\mathbf{e}_{1}\times \mathbf{e}_{2}}{e},
\end{equation*}%
where $e\ $is the determinant of $e_{(A)}^{i}$. At the centre line, we have 
\begin{equation}
\hat{n}=-\sin u\cos {\frac{u}{2}}\hat{\imath}-\cos u\cos {\frac{u}{2}}\hat{%
\jmath}+\sin {\frac{u}{2}}\hat{k}.  \tag{4}
\end{equation}%
Note that if $u=0$ we find $\hat{n}=-\hat{\jmath}$ and if $u=2\pi $ we get $%
\hat{n}=\hat{\jmath}$. While if $u=4\pi $ we obtain $\hat{n}=-\hat{\jmath}$
as expected.

\bigskip \ 

\noindent \textbf{3. Brief review of constraint Hamiltonian system}

\smallskip \ 

Let us first recall the traditional steps to quantize a classical system.
One first consider the action:

\begin{equation}
S[q]=\int dtL(q,\frac{dq}{dt}),  \tag{5}
\end{equation}%
where the Lagrangian $L$ is a function of the $q^{i}$-coordinates and the
corresponding velocities $dq^{i}/dt,$ with $i=1,\dots ,n$.

Defining the canonical momentum $p_{i}$ conjugate to $q^{i}$ by%
\begin{equation}
p_{i}\equiv \frac{\partial L}{\partial (\frac{dq^{i}}{dt})},  \tag{6}
\end{equation}%
it allow us to rewrites the action (1) in the form

\begin{equation}
S[q,p]=\int dt(\frac{dq^{i}}{dt}p_{i}-H_{c}),  \tag{7}
\end{equation}%
where $H_{c}=H_{c}(q,p)$ is the canonical Hamiltonian,

\begin{equation}
H_{c}(q,p)\equiv \frac{dq^{i}}{dt}p_{i}-L.  \tag{8}
\end{equation}

The Poisson bracket, for arbitrary functions $f(q,p)$ and $g(q,p)$ of the
canonical variables $q$ and $p$ is defined as

\begin{equation}
\{f,g\}=\frac{\partial f}{\partial q^{i}}\frac{\partial g}{\partial p_{i}}-%
\frac{\partial f}{\partial p_{i}}\frac{\partial g}{\partial q^{i}}.  \tag{9}
\end{equation}%
Using (5) we find that the basic Poisson brackets are

\begin{equation}
\{q^{i},q^{j}\}=0,  \tag{10}
\end{equation}

\begin{equation}
\{p_{i},p_{j}\}=0,  \tag{11}
\end{equation}

\begin{equation}
\{q^{i},p_{j}\}=\delta _{j}^{i}.  \tag{12}
\end{equation}%
Here, the symbol $\delta _{j}^{i}$ denotes a Kronecker delta.

The transition to quantum mechanics is made by promoting the Hamiltonian $%
H_{c}(q,p)$ as an operator $\hat{H}_{c}(\hat{q},\hat{p})$ via the commutators

\begin{equation}
\lbrack \hat{q}^{i},\hat{q}^{j}]=0,  \tag{13}
\end{equation}

\begin{equation}
\lbrack \hat{p}^{i},\hat{p}^{j}]=0,  \tag{14}
\end{equation}

\begin{equation}
\lbrack \hat{q}^{i},\hat{p}^{j}]=i\delta ^{ij},  \tag{15}
\end{equation}%
with $\hbar =1$. Here, $[\hat{A},\hat{B}]=\hat{A}\hat{B}-\hat{B}\hat{A}$
denotes a commutator for any arbitrary operators $\hat{A}$ and $\hat{B}$.
Note that the algebra (13)-(15) is obtained from (10)-(12) via the
transition $\{f,g\} \rightarrow \frac{1}{i}[\hat{f},\hat{g}]$ for two
arbitrary functions $f$ and $g$ in the phase space. In addition, one assumes
the quantum formula [17-18]

\begin{equation}
\hat{H}_{c}\mid \Psi >=i\frac{\partial }{\partial t}\mid \Psi >,  \tag{16}
\end{equation}%
which determines the physical states $\mid \Psi >$ (see Refs. [12] and [13]
for details).

It is worth mentioning that (16) can be obtained from a classical constraint
system. In fact, let us assume that the Lagrangian $L(q,\frac{dq}{dt})$ has
the form%
\begin{equation}
L=\frac{1}{2}m_{0}(\frac{dq}{dt})^{2}-V(q).  \tag{17}
\end{equation}%
In this case the action (1) becomes.

\begin{equation}
S[q]=\int dt(\frac{1}{2}m_{0}(\frac{dq}{dt})^{2}-V(q)).  \tag{18}
\end{equation}%
If one introduces an arbitrary parameter $\tau $ such that $q^{0}\equiv
t=t(\tau )$ one find that (18) can be rewritten as

\begin{equation}
S[q^{0},q^{i}]=\int d\tau (\frac{1}{2}m_{0}(\dot{q}^{0})^{-1}\dot{q}^{2}-%
\dot{q}^{0}V(q)),  \tag{19}
\end{equation}%
where $\dot{A}\equiv \frac{dA}{d\tau }$ for an arbitrary variable $A$. The
expression (19) suggests a possible redefininition of the Lagrangian as

\begin{equation}
\mathcal{L}=\frac{1}{2}m_{0}(\dot{q}^{0})^{-1}\dot{q}^{2}-\dot{q}^{0}V(q) 
\tag{20}
\end{equation}%
Thus, in addition to the definition of the momentum $p_{i}$ given by

\begin{equation}
p_{i}\equiv \frac{\partial \mathcal{L}}{\partial (\dot{q}^{i})},  \tag{21}
\end{equation}%
one may also consider the momentum%
\begin{equation}
p_{0}\equiv \frac{\partial \mathcal{L}}{\partial (\dot{q}^{0})},  \tag{22}
\end{equation}

Using (20) one sees that (21) and (22) become%
\begin{equation}
p^{i}\equiv m_{0}(\dot{q}^{0})^{-1}\dot{q}^{i},  \tag{23}
\end{equation}%
and%
\begin{equation}
p_{0}\equiv -\frac{1}{2}m_{0}(\dot{q}^{0})^{-2}\dot{q}^{2}-V(q),  \tag{24}
\end{equation}%
respectively. These two expressions can be combined to obtain the first
class constraint

\begin{equation}
p_{0}+\frac{p^{i}p_{i}}{2m_{0}}+V(q)=0.  \tag{25}
\end{equation}%
We can also show using (23) and (24) that in this case $H_{c}$ vanishes
identically.

Thus, the action functional associated with $\mathcal{L}$ can be written in
the phase space as

\begin{equation}
\mathcal{S}[q^{0},q^{i},p_{0},p_{j},\lambda ^{0}]=\int d\tau (\dot{q}%
^{0}p_{0}+\dot{q}^{i}p_{i}-\lambda ^{0}(p_{0}+\frac{p^{i}p_{i}}{2m_{0}}%
+V(q)),  \tag{26}
\end{equation}%
where $\lambda ^{0}$ is a Lagrange multipliers. Defining

\begin{equation}
H_{0}\equiv p_{0}+\frac{p^{i}p_{i}}{2m_{0}}+V(q),  \tag{27}
\end{equation}%
we can rewrite (26) as

\begin{equation}
\mathcal{S}=\int d\tau (\dot{q}^{0}p_{0}+\dot{q}^{i}p_{i}-\lambda ^{0}H_{0}).
\tag{28}
\end{equation}

At the quantum level we still have the algebra (13)-(15), but instead of
(16) one imposes the condition

\begin{equation}
\hat{H}_{0}\mid \Psi >=0,  \tag{29}
\end{equation}%
on the physical states $\mid \Psi >$. Observe that (29) leads to (16) if we
write $\hat{p}_{0}=-i\frac{\partial }{\partial t}$ and $\hat{H}_{c}=\frac{%
\hat{p}^{i}\hat{p}_{i}}{2m_{0}}+V(q)$. In other words, the Schr\"{o}dinger
equation (16) can be derived from the constraint structure (29).

In general if in addition to $H_{0}$ one introduces additional first class
constrains $H_{m}(q,p)\approx 0$, with $m=1,2...,n$, then (28) can be
extended to the form

\begin{equation}
S[q,p]=\int_{t_{i}}^{t_{f}}d\tau (\dot{q}^{0}p_{0}+\dot{q}^{i}p_{i}-\lambda
^{0}H_{0}-\lambda ^{m}H_{m}),  \tag{30}
\end{equation}%
and the physical states $\mid \Psi >$ must now satisfy the conditions

\begin{equation}
\hat{H}_{0}\mid \Psi >=0,  \tag{31}
\end{equation}%
and

\begin{equation}
\hat{H}_{m}\mid \Psi >=0.  \tag{32}
\end{equation}

\bigskip \ 

\noindent \textbf{4. Spinning particle in a M\"{o}bius strip}

\smallskip \ 

Our spinning object model is in a sense inspired in the relativistic
spherical top theory (see Ref. [13] and references therein). However, we
shall focus here in the nonrelativistic case. Specifically, let us describe
the motion of a non relativistic spinning object with the variables $%
x^{i}(\tau )$ and $e_{i}^{(a)}(\tau )$, where $\tau $ is an arbitrary
parameter; $x^{i}(\tau )$ determines the position of the system, while $%
e_{i}^{(a)}(\tau )$ is an orthonormal frame used to determine its
orientation. We also introduce the canonical conjugate variables $p_{i}$ and 
$\Sigma _{ij}$ associated with $x^{i}$and $e_{i}^{(a)}$ respectively. The
orthonormal character of the variables $e_{i}^{(a)}$ can be expressed by

\begin{equation}
e_{i}^{(a)}e_{j}^{(b)}\delta _{(ab)}=\delta _{ij}.  \tag{33}
\end{equation}%
Therefore the angular velocity

\begin{equation}
\sigma _{ij}=e_{i}^{(a)}\frac{d}{d\tau }e_{j}^{(b)}\delta _{(ab)},  \tag{34}
\end{equation}%
becomes antisymmetric tensor, that is, $\sigma _{ij}=-\sigma _{ji}$. Thus, $%
\sigma _{ij}$ has only three rotational degrees of freedom, as it is
required.

Our main assumption is that such a spinning object is confined to move along
the meridian (central line) of a M\"{o}bius strip (see Figure 2).

\begin{figure}[htbp]
\centering
\includegraphics[scale=0.45,keepaspectratio=true]{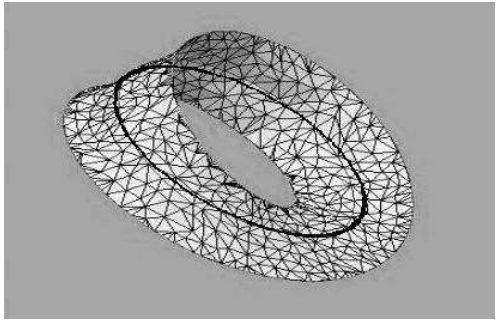}
\caption{{\protect \footnotesize A M\"{o}bius strip showing its meridian
cycle }}
\end{figure}

Specifically, we shall consider a theory in which the motion of a spinning
object is determined by the action

\begin{equation}
\begin{array}{c}
S=\int d\tau \lbrack \dot{x}^{0}p_{0}+\dot{x}^{i}p_{i}+\frac{1}{2}\sigma
^{ij}\Sigma _{ij}-\lambda (p_{0}+\frac{p^{i}p_{i}}{2m_{0}}+\frac{\Sigma
^{ij}\Sigma _{ij}}{2m_{0}\rho ^{2}}+V(x^{i})) \\ 
\\ 
-\xi (x^{i}x_{i}-r^{2})-\eta _{i}(\frac{1}{2}\varepsilon ^{ijk}\Sigma
_{jk}-sp^{i})-\kappa _{(a)}^{i}(e_{i}^{(a)}(\tau )-e_{i}^{(a)}(\tau +4\pi
k)].%
\end{array}
\tag{35}
\end{equation}%
Here, $\lambda ,$ $\xi ,$ $\eta $ and $\kappa $are Lagrange multipliers, $%
\rho $ is a constant describing the size of the spinning object, and finally 
$r^{2}$ and $s$ are considered to be constants of the motion.

Varying the parameters $\lambda ,$ $\xi ,$ $\eta $ and $\kappa $ we obtain
the constraints

\begin{equation}
p_{0}+\frac{p^{i}p_{i}}{2m_{0}}+\frac{\Sigma ^{ij}\Sigma _{ij}}{2m_{0}\rho
^{2}}+V=0,  \tag{36}
\end{equation}%
\begin{equation}
x^{i}x_{i}-r^{2}=0,  \tag{37}
\end{equation}%
\begin{equation}
\varepsilon ^{ijk}\Sigma _{jk}-sp^{i}=0,  \tag{38}
\end{equation}%
and%
\begin{equation}
e_{i}^{(a)}(\tau )-e_{i}^{(a)}(\tau +4\pi n)=0.  \tag{39}
\end{equation}%
The first constraint determines the relation between energy $p_{0}$, kinetic
energy $\frac{p^{i}p_{i}}{2m_{0}}$, rotational energy $\frac{\Sigma
^{ij}\Sigma _{ij}}{2m_{0}\varrho ^{2}}$ and potential energy $V$. The
constraint (37) assure that the object moves on a circle. While the
constraints (38) and (39) establishes the rotational motion of an
orthonormal frame attached to the M\"{o}bius strip. The constraint (38) is
used to force the system to have parallel angular momentum to the linear
momentum during the motion along the meridian of the M\"{o}bius strip. Note
that (39) specifies that the orthonormal frame needs to rotate twice in
order to return to the original orientation.

Solving (38) and substituting the result into (36) gives

\begin{equation}
p_{0}+\frac{p^{i}p_{i}}{2m_{0}}+\frac{s^{2}p^{i}p_{i}}{2m_{0}\rho ^{2}}+V=0,
\tag{40}
\end{equation}%
which can be rewritten as%
\begin{equation}
p_{0}+\frac{(1+\frac{s^{2}}{\rho ^{2}})p^{i}p_{i}}{2m_{0}}+V=0.  \tag{41}
\end{equation}%
Therefore, by quantizing (41) we find the Sch\"{o}dinger-like equation

\begin{equation}
\left( \hat{p}_{0}+\frac{(1+\frac{s^{2}}{\rho ^{2}})\hat{p}^{i}\hat{p}_{i}}{%
2m_{0}}+V\right) \left \vert \psi \right \rangle =0,  \tag{42}
\end{equation}%
which in a coordinate representation can be written as%
\begin{equation}
\left( -i\hslash \frac{\partial }{\partial t}-\frac{\hslash ^{2}(1+\frac{%
s^{2}}{\rho ^{2}})\nabla ^{2}}{2m_{0}}+V\right) \psi (t,x^{i})=0.  \tag{43}
\end{equation}

Considering that the system moves in a circle determined by the constraint
(37) we get%
\begin{equation}
\left( -i\hslash \frac{\partial }{\partial t}-\frac{\hslash ^{2}}{%
2m_{0}^{\prime }}\frac{d^{2}}{d\theta ^{2}}+V\right) \psi (t,\theta )=0, 
\tag{44}
\end{equation}%
where%
\begin{equation}
m_{0}^{\prime }=\frac{m_{0}}{(1+\frac{s^{2}}{\rho ^{2}})}.  \tag{45}
\end{equation}%
The formula (44) is the key equation for our approach. It is important to
remark that the periodicity in the orthonormal frame (39) translates at the
level of the quantum equation (44) in the requirement $\psi (t,\theta )=\psi
(t,\theta +4n\pi )$.

We shall now study some of the consequences of the key equation (44) in
several cases. This corresponds to a number of specific forms of the
potential $V$.

\smallskip \ 

\noindent \textbf{Example 1}:

\smallskip \ 

The simplest example is a free system, with $V=0$. In this case (44) is
reduced to

\begin{equation}
\left( -i\hslash \frac{\partial }{\partial t}-\frac{\hslash ^{2}}{%
2m_{0}^{\prime }}\frac{d^{2}}{d\theta ^{2}}\right) \psi (t,\theta )=0. 
\tag{46}
\end{equation}%
In the stationary case (46) becomes

\begin{equation}
\left( -E-\frac{\hslash ^{2}}{2m_{0}^{\prime }}\frac{d^{2}}{d\theta ^{2}}%
\right) \psi (\theta )=0.  \tag{47}
\end{equation}%
Thus, in this case we have the general solution 
\begin{equation}
\psi (\theta )=\sum_{n=1}^{\infty }(a_{n}e^{\frac{in\theta p_{\theta }}{%
2\hslash }}-a_{n}^{\ast }e^{-\frac{in\theta p_{\theta }}{2\hslash }}), 
\tag{48}
\end{equation}%
where $a_{n}$ is a complex constant. Note that $\psi (\theta )=\psi (\theta +%
\frac{4n\pi }{p_{\theta }})$ as expected. Therefore we find that the energy
eigenvalues are given by%
\begin{equation}
E_{n}=\frac{n^{2}p_{\theta }^{2}}{8m_{0}^{\prime }}.  \tag{49}
\end{equation}

Consider a proton moving along the meridian of a M\"{o}bius strip. The
root-mean square charge radius $r_{p}$ and the mass $m_{0}$ of the proton
can be taken as [19-20]

\begin{equation}
\rho \equiv r_{p}=0.8418\times 10^{-15}m,  \tag{50}
\end{equation}%
and%
\begin{equation}
m_{0}=1.673\times 10^{-27}\quad Kg,  \tag{51}
\end{equation}%
respectively. While, since the proton spin is $\Sigma =\frac{\hslash }{2}$,
we can assume that the quantity $s=\frac{\Sigma }{p}$ is also of the order

\begin{equation}
s\sim 10^{-15}m.  \tag{52}
\end{equation}%
Therefore, from (49) we find the ground state energy is or the order%
\begin{equation}
E_{1}\sim \frac{p_{\theta }^{2}}{5.6\times 10^{-27}\quad Kg}.  \tag{53}
\end{equation}

\smallskip \ 

\noindent \textbf{Example 2}:

\smallskip \ 

As a second example let us now assume that

\begin{equation}
V=-\frac{e^{2}}{r}.  \tag{54}
\end{equation}%
We will look for a solutions of the equation

\begin{equation}
\left( \frac{-\hslash ^{2}\nabla ^{2}}{2m_{0}^{\prime }}-\frac{e^{2}}{r}%
\right) \psi (x^{i})=E\psi (x^{i}).  \tag{55}
\end{equation}%
Observe that (50) can be interpreted as the Schr\"{o}dinger equation for the
Hydrogen atom. However, one should keep in mind that we are choosing
circular motion for the electron and that the wave function $\psi (x^{i})$
has periodicity $4\pi $. In fact in this case (50) can be reduced to

\begin{equation}
\frac{-\hslash ^{2}}{2m_{0}^{\prime }}(\frac{d^{2}}{dr^{2}}+\frac{2d}{rdr}+%
\frac{d^{2}}{r^{2}d\theta ^{2}})-\frac{e^{2}}{r})\psi (r,\theta )=E\psi
(r,\theta ).  \tag{56}
\end{equation}

We can propose the solution

\begin{equation}
\psi (r,\theta )=R(r)e^{i\frac{k\theta }{2}},  \tag{57}
\end{equation}%
with eigenvalue equation

\begin{equation}
\hat{p}_{\theta }^{2}\psi (r,\theta )=-\hslash ^{2}\frac{d^{2}}{d\theta ^{2}}%
\psi (r,\theta )=\frac{1}{4}\hslash ^{2}k^{2}\psi (r,\theta ).  \tag{58}
\end{equation}%
Here $k$ is considered to be an integer. Using (57) the formula (56) becomes

\begin{equation}
\left( \frac{-\hslash ^{2}}{2m_{0}^{\prime }}(\frac{d^{2}}{dr^{2}}+\frac{2d}{%
rdr})+\frac{\hslash ^{2}k^{2}}{8m_{0}^{\prime }r^{2}}-\frac{e^{2}}{r}\right)
R(r)=ER(r).  \tag{59}
\end{equation}%
Now, following standard procedure we find that the energy eigenvalues are

\begin{equation}
E_{n}=-\frac{1}{2}m_{0}^{\prime }c^{2}\frac{\alpha ^{2}}{n^{2}},  \tag{60}
\end{equation}%
where%
\begin{equation}
\alpha =\frac{e^{2}}{\hslash c},  \tag{61}
\end{equation}%
is the fine structure constant and $n$ can take the values $1,2,3,...$. But
with with the restriction 
\begin{equation}
n^{2}-n\geq \frac{k^{2}}{4},  \tag{62}
\end{equation}%
instead of the usual one 
\begin{equation}
n^{2}-n\geq l(l+1).  \tag{63}
\end{equation}%
Thus, although our result (60) looks as the normal energy eigenvalues of the
Hydrogen atom in fact it differs in two fundamental aspects: (1) according
to (45) the constant $m_{0}^{\prime }$ depends of the size $\rho $ and the
the parameter $s$ associated with the spin of the rotating object
respectively, (2) the positive integer $n$ must satisfy (62) instead of (63).

\smallskip \ 

\noindent \textbf{Example 3}:

\smallskip \ 

As our final example we shall consider a constant magnetic flux $A$ through
the meridian ring of the M\"{o}bius strip. In this case, one can show that
(44) leads to%
\begin{equation}
\left( -E-\frac{\hslash ^{2}}{2m_{0}^{\prime }}(\frac{d}{d\theta }%
+A)^{2}\right) \psi (\theta )=0.  \tag{64}
\end{equation}%
The eigenfunction solution of (64) can be written as

\begin{equation}
\psi (\theta )=e^{\frac{in\theta }{2}}.  \tag{65}
\end{equation}%
This yields the energy eigenvalues

\begin{equation}
E_{n}=\frac{\hslash ^{2}}{2m_{0}^{\prime }}(\frac{n}{4}-A)^{2},  \tag{66}
\end{equation}%
where $n=0,\pm 1,\pm 2,...$. It is interesting to mention that classically
the constant $A$ can be be associated with a topological term and
consequently it is not observable. But (60) shows that at the quantum level $%
A$ changes the energy eigenvalues. However, since $A$ is related to
topological structure it may be combined with the boundary conditions which
in turn are related to the topology of the M\"{o}bius strip. This
observation clearly motives further studies on the subject.

\bigskip \ 

\smallskip \ 

\noindent \textbf{5. Final remarks}

\smallskip \ 

In this work we have derived the Schr\"{o}dinger-like equation for a
spinning particle moving in the meridian of a M\"{o}bius strip. The main
advantage of our Lagrangian approach is that important mathematical tools of
the constrains Hamiltonian theory can be used. In particular, one can
determine in a systematic way the symmetries of the theory.

We show the advances of our method considering three examples; (1) $V=0$ and
(2) $V=-\frac{e^{2}}{r}$, and $p\rightarrow p-A$. In the three cases the
functions states have the required boundary conditions and the energy
eigenvalues are quantized. In the first case the energy eigenvalues $E_{n}$
goes as as $n^{2}$. While in the second case the $E_{n}$ goes as $\frac{1}{%
n^{2}}$, having the form of the hydrogen atom but with the mass constant $%
m_{0}^{\prime }$ depending on the size $\rho $ and the internal parameter $s$
associated with the spin of the rotating object. Furthermore, the positive
integer $n$ is restricted to satisfy (62) instead of (63).

It may be interesting for further research to consider the motion of
spinning system in the whole M\"{o}bius strip either at the classical or
quantum level. In particular, it is worth pursuing what the eigenfunctions
and eigenvalues would be like for such system.

\begin{center}
\bigskip \ 

\bigskip \ 

\smallskip \ 

\textbf{Acknowledgments}
\end{center}

J. A. Nieto would like to thank the Departamento de Investigacion en F\'{\i}%
sica for the hospitality. This work was partially supported by PROFAPI-UAS
2009.

\begin{center}
\smallskip \ 
\end{center}

\end{document}